\documentclass[runningheads]{llncs}
\usepackage[binary-units,per-mode=symbol]{siunitx}
\usepackage{hyperref}
\usepackage{booktabs}
\usepackage[table]{xcolor}  
\usepackage[utf8]{inputenc}
\usepackage{url}
\usepackage{microtype}
\usepackage{subfig,subfloat}
\usepackage{tikz}
\usepackage{tikzscale}
\usetikzlibrary{external}
\usepackage{pgfplots}
\usepgfplotslibrary{external}

\usepackage{dcolumn}
\usepackage{algorithm}
\usepackage{algpseudocode}
\usepackage[shortcuts]{extdash}
\newcolumntype{g}{D{.}{.}{-1}}

\definecolor{bblue}{HTML}{4F81BD}
\definecolor{rred}{HTML}{C0504D}
\definecolor{ggreen}{HTML}{9BBB59}
\definecolor{ppurple}{HTML}{9F4C7C}

\newcommand*{\codepoint}[1]{\texttt{U+\MakeUppercase{#1}}}
\newcommand*{\codepointrange}[2]{\texttt{U+\MakeUppercase{#1}\ldots\MakeUppercase{#2}}}
\newcommand*{\hexrange}[2]{\texttt{0x\MakeUppercase{#1}\ldots\MakeUppercase{#2}}}

\newcommand*{\restartrowcolors}{%
  \ifhmode\unskip\fi
  \vadjust{%
    \global\rownum=0 %
  }%
}

\usepackage{graphicx}
\begin{document}
\title{Unicode at Gigabytes per Second}
\author{Daniel Lemire}%\orcidID{0000-0003-3306-6922}
\authorrunning{D. Lemire}
\institute{DOT-Lab Research Center, University of Quebec (TELUQ), Montreal, Canada\\
\email{daniel.lemire@teluq.ca}}
\maketitle              % typeset the header of the contribution
\begin{abstract}
We often represent text using Unicode formats (UTF-8 and UTF-16).  The UTF-8 format is increasingly popular, especially on the web (XML, HTML, JSON, Rust, Go, Swift, Ruby). The UTF-16 format is most common in Java, .NET, and inside operating systems such as Windows.

Software systems frequently have to convert text from one Unicode format to the other. While recent disks have bandwidths of \SI{5}{\gibi\byte\per\second} or more, conventional approaches transcode non-ASCII text at a fraction of a gigabyte per second.

We show that we can validate and transcode Unicode text at gigabytes per second on current systems (x64 and ARM) without sacrificing safety. Our open-source library can be ten times faster than the popular ICU library on non-ASCII strings and even faster on ASCII strings.

%Deadline is July 28th.

\end{abstract}
\section{Introduction}
\let\thefootnote\relax\footnotetext{This  manuscript is based on a forthcoming long-form article written with Wojciech Muła, \emph{Transcoding Billions of Unicode Characters per Second with SIMD Instructions}.}
From the early days of computing, programmers have had to represent characters in software. They needed to agree to standards so that different software written by different vendors would be interoperable. One of the earliest such standards is ASCII---first specified in the early 1960s. The ASCII standard is still popular today: it uses one byte per character---with the most significant bit set to zero. Unfortunately, ASCII could only ever represent up to 128~characters---far less than needed. 

Thus many diverging standards emerged for representing characters in software. The existence of multiple incompatible formats made the production of interoperable localized software difficult. Conversion between some formats could sometimes be lossy or ambiguous.

Unicode arose in the late 1980s as an attempt to provide a single agreed-upon standard. Initially, it was believed that using 16~bits per character would be sufficient. However, engineers realized over time that a wider range of characters should be supported---if the standard was to be universal.
Thus the Unicode standard was extended to potentially include up to \num{1114112}~characters. Characters are sometimes called code points and represented as  integer values between  0 and \num{1114112}. In practice, only a small fraction of all possible code points have been assigned, but more are assigned over time with each Unicode revision.
The Unicode standard is an extension of the ASCII standard in the sense that the first 128~Unicode code-point values match the ASCII characters.

There are several ways to represent Unicode characters in bytes. Due to the original expectation that Unicode would fit in 16-bit space, a format based on 16-bit words (UTF-16) format was published in 1996 and  formalized in 2000~\cite{rfc2781}. It may use either 16-bit or 32-bit per character. 
The UTF-16 format was adopted by programming languages such as  Java, and became a default under Windows.

Unfortunately, UTF-16 is not backward compatible with ASCII at a byte level. Thus an ASCII compatible format was proposed and formalized in 2003: UTF-8~\cite{rfc3629}. Over time, it became widely used. Text interchange formats such as JSON, HTML or XML are expected to be in the UTF-8 format. Programming languages such as Go, Rust and Swift use UTF-8 by default. When used as part of data interchange documents, UTF-8 is commonly more concise due to its ability to use  one~byte per character to represent ASCII text.

Though UTF-8 dominates in many applications, it does not make UTF-16 obsolete. The UTF-16 format has advantages. Indeed, most text represented in the UTF-16 format has exactly 2-byte per character, except for the occasional special character (e.g., an emoji). Having a flat 2-byte per character makes some operations faster. 
Both formats require validation: not all arrays of bytes are valid. However, the UTF-8 format is more expensive to validate. In some cases, UTF-16 may be even more concise as well (e.g., when representing Chinese text).

Thus, for the foreseeable future, we need to validate both UTF-16 and UTF-8 strings, and to convert (transcode) text between the two formats.

We should expect these operations to be fast, and they are. However, speed and efficiency are relative.
Cloud vendors offer high bandwidth  between node instances  (e.g., \SI{3.3}{\gibi\byte\per\second}) and the bandwidth of disks is rising fast (e.g., \SI{5}{\gibi\byte\per\second} with PCIe~4.0)~\cite{bar2008}. We believe that we should be
able to match such high speeds inside the processor when transcoding text.

For fast processing, we should seek to make the best possible use out of our processors.
Commodity processors support
single-instruction-multiple-data (SIMD) instructions. These instructions operate on several words at once unlike regular instructions. 
Starting with the Haswell microarchitecture (2013), Intel and AMD processors support the AVX2 instruction set and 256-bit vector registers. Most mobile phones tablets have 64-bit ARM processors (\texttt{aarch64}) with NEON instructions (128-bit registers).
Hence, on recent x64 processors, we can compare two strings of 32~characters in a single instruction. Algorithms designed for SIMD instructions typically require fewer instructions per byte. Software that retires fewer instructions tends to use less power and to be faster.

\section{Related Work}
\label{sec:relatedwork}

There are many ways to validate and transcode Unicode text. One may use series of branches or finite-state approaches~\cite{Hoehrmann}. We are most interested in the fastest techniques.
Keiser and Lemire~\cite{keiser2020validating} describe a fast  SIMD-based UTF-8 validation directly on byte streams. We make use of their approach (see \S~\ref{sec:algorithms}).
 
Cameron~\cite{cameron2008case} proposed that we transform text using \emph{bit streams}. Given byte arrays, the bit-stream approach creates eight bit arrays, each one corresponding to a bit position within a byte. There is one bit stream for the least significant bit, one for the second least significant bit, and so forth. We must first transform the input data into such bit streams and then convert the data back from bit streams to an array of bytes. Cameron applied this strategy to  UTF-8 to UTF-16 transcoding.

Independently, Inoue et al.~\cite{Inoue2008} proposed a UTF-8 to UTF-16 SIMD-accelerated transcoder, but it  does not validate its inputs. The authors did not make their implementation (for PowerPC processors) available.

We are not aware of any further work in the scientific literature regarding the application of SIMD instruction to the validation or transcoding of Unicode text. Except  for fast ASCII paths, we do not know of any widespread use of SIMD instructions for validating or transcoding Unicode text.

\section{The Unicode Formats}

 ASCII characters require one byte with UTF-8 and two bytes with UTF-16. UTF-16 can represent all characters---except for the supplemental characters such as emojis---using two bytes. The UTF-8 format uses two bytes for Latin, Hebrew and Arabic alphabets. Asiatic characters (including Chinese and Japanese) require three UTF-8 bytes. Both UTF-8 and UTF-16 require 4~bytes for the supplemental characters. % There is only one way to encode a given Unicode character.
 We often represent Unicode characters using its integer value in hexadecimal as, for example, \codepoint{7F} (for 127).
 
UTF-8 encodes values in sequences of one to four~bytes. We refer to the first byte of a sequence as a  leading byte; the most significant bits of the leading byte indicate the length of the sequence:
\begin{itemize}
\item If the most significant bit is zero, we have a sequence of one byte (ASCII).
\item If the three most significant bits are \texttt{110}, we have a two-byte sequence.
\item If the four most significant bits are \texttt{1110}, we have a three-byte sequence.
\item Finally, if the five most significant bits are \texttt{11110}, we have a four-byte sequence.
\end{itemize}
All bytes following the leading byte in a sequence are continuation bytes, and they must have their two most significant bits as \texttt{10}.
Except for the required most significant bit sequences, other bits (from 7~bits to 21~bits) provide the code-point value. The most significant bits of the code-point value are in the leading byte, followed by lesser significant bits in the second byte and so forth, with the least significant bits in the last byte of the sequence. Valid UTF-8 sequences must follow the following exhaustive rules:
\begin{enumerate}
\item The five most significant bits of any byte cannot be all ones.
\item The leading byte must be followed by the right number of continuation bytes. 
\item A continuation byte must be preceded by a leading byte.
\item The decoded character must be larger than \codepoint{7F} for two-byte sequences, larger than \codepoint{7FF} for three-byte sequences, and larger than \codepoint{FFFF} for four-byte sequences. 
\item The decoded code-point value  must be less  than \num{1114112}. 
\item The  code-point value must not be in the range \codepointrange{D800}{DFFF}. 
%\end{tabular}
\end{enumerate}

In the UTF-16 format, characters in 
\codepointrange{0000}{D7FF} and \codepointrange{E000}{FFFF} are stored as 16-bit values---using two bytes. The characters in the range
\codepointrange{010000}{10FFFF} require two 16-bit words (a surrogate pair). The first word in the pair is in \hexrange{d800}{dbff} whereas the second word is in \hexrange{dc00}{dfff}. The code-point value is made of the  10~least significant bits of the two words---using the second word as least significant---adding \texttt{0x10000} to the result. During validation, only the possible surrogate pairs require attention.

\section{Algorithms}
\label{sec:algorithms}

All commodity software with SIMD instructions (e.g., x64, ARM, POWER) have fast instructions to  permute bytes within a SIMD register  according to a sequence of indexes. Our transcoding techniques depend critically on this feature: we code in a table the necessary parameters---including the indexes (sometimes called shuffle masks)---necessary to process a variety of incoming characters.

Our  accelerated  UTF-8 to UTF-16 transcoding algorithm  processes  up to 12~input UTF-8 bytes at a time. From the input bytes, we can quickly determine the leading bytes and thus the end of each character. We use a 12-bit word as a key in a 1024-entry table. Each entry in the table contains the number of UTF-8 bytes that will be consumed and an index into another table where we find shuffle masks. The value of the index into the other table also determines one of three possible code paths. The first 64~index values  indicate  that we have 6~characters spanning between one and two bytes.  Index values in $[64,145)$ indicate that we have 4~characters spanning between one and three bytes. The remaining indexes represent the general case: 3~characters spanning between one and four bytes.
The shuffle mask can then be applied to the 12~input bytes to form a vector register that can be transformed efficiently.
We use this 12-byte routine inside  64-byte blocks. After loading a 64-byte block,  
we apply the Keiser-Lemire validation routine~\cite{keiser2020validating}. Afterward,
we identify the leading bytes, and then process the block in multiple iterations, using up to 12~bytes each time. In the special case where all 64~bytes are ASCII, we use a fast path.
For even greater efficiency, we have three other fast paths within the 12-byte routine: we check whether the next 16~bytes are ASCII bytes,  whether they are all two-byte characters, or all  three-byte characters.

Our UTF-16 to UTF-8 algorithm iteratively reads a block of input bytes in a SIMD register.  If all 16-bit words in the loaded SIMD register are in the range \codepointrange{0000}{007f}, we use a fast routine to convert the 16~input bytes into eight equivalent ASCII bytes.
 If all 16-bit words are in the range \codepointrange{0000}{07ff}, then we use a fast routine to produce sequences of one-byte or two-byte UTF-8 characters. Given an 8-bit bitset which indicates which 16-bit words are ASCII, we load a byte value from a table indicating how many bytes will be written, and a 16-byte shuffle mask. 
If all 16-bit words are in the ranges \codepointrange{0000}{d7ff}, \codepointrange{e000}{ffff}, we use another similar specialized  routine to produce sequences of one-byte, two-byte and three-byte UTF-8 characters.
Otherwise, when we detect that the input register contains at least one part of a surrogate pair, we fall back to a conventional code path.

\section{Experiments}
\label{sec:experiments}

\begin{figure}
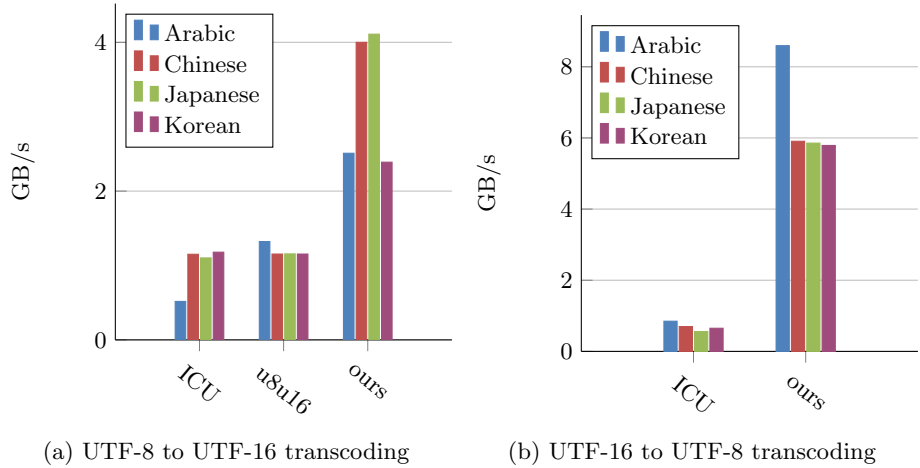
\centering
\subfloat[UTF-8 to UTF-16 transcoding]{
\includegraphics[width=0.49\textwidth]{lipsumspeedutf8utf16.tikz}
}
\subfloat[UTF-16 to UTF-8 transcoding]{
\includegraphics[width=0.49\textwidth]{lipsumspeedutf16utf8.tikz}
}

\caption{\label{fig:transcoding} Transcoding speeds for various test files. }
\end{figure}

We make available our software as a portable open-source  C++ library.\footnote{\url{https://github.com/simdutf/simdutf}}
As a benchmarking system, we use a recent AMD processor (AMD EPYC 7262,  Zen~2 microarchitecture, \SI{3.39}{\GHz}) and GCC~10.
We compare against a popular library: International Components for Unicode (ICU)~\cite{icu} (version~67.1).
We also use the \texttt{u8u16} library~\cite{cameron2008case}. Unlike ICU and our own work, the  \texttt{u8u16} library only provides UTF-8 to UTF-16 transcoding.
For our experiments, we use lipsum text in various languages.\footnote{\url{https://github.com/rusticstuff/simdutf8}} 
All of our transcoding tests include validation.
To measure the speed, we record the time by repeating the task \num{2000}~times. We compare the average time with the minimal time and find that we have an accuracy of at least 1\%. We divide the input volume by the time required for the transcoding. Fig.~\ref{fig:transcoding} shows our results. 
Our UTF-8 to UTF-16 transcoding speed exceeds  \SI{4}{\gibi\byte\per\second} for Chinese and Japanese texts, which is about four times faster than ICU. In our tests, the \texttt{u8u16} library only surpasses ICU significantly for Arabic.
Our UTF-16 to UTF-8 transcoding speed is nearly \SI{6}{\gibi\byte\per\second} in all tests which is nearly ten~times faster than ICU.

For ASCII transcoding (not shown in the figures), we achieve  \SI{36}{\gibi\byte\per\second} for UTF-16 to UTF-8 transcoding, and  \SI{20}{\gibi\byte\per\second} for UTF-8 to UTF-16 transcoding. Effectively, we are so fast that we are nearly limited by memory bandwidth. Comparatively, ICU delivers \SI{2}{\gibi\byte\per\second} and \SI{1}{\gibi\byte\per\second} in our tests.

\section{Conclusion}

Our SIMD-based transcoders can surpass popular transcoders (e.g., ICU) by a wide margin (e.g., $4\times$). Our UTF-16 to UTF-8 transcoder achieves speed of about \SI{6}{\gibi\byte\per\second} for many Asiatic languages using a recent x64 processor. In some cases, we achieve  \SI{4}{\gibi\byte\per\second}  for UTF-8 to UTF-16 transcoding with full validation. For ASCII inputs, we achieve tens of gigabytes per second.

%{rfc2781}

\bibliographystyle{splncs04}
\bibliography{spireunicodelemire}

\begin{thebibliography}{1}
\providecommand{\url}[1]{\texttt{#1}}
\providecommand{\urlprefix}{URL }
\providecommand{\doi}[1]{https://doi.org/#1}

\bibitem{bar2008}
Barr, J.: {The Floodgates Are Open – Increased Network Bandwidth for EC2
  Instances}.
  \url{https://aws.amazon.com/blogs/aws/the-floodgates-are-open-increased-network-bandwidth-for-ec2-instances/}
  [last checked July 2021] (2018)

\bibitem{cameron2008case}
Cameron, R.D.: A case study in {SIMD} text processing with parallel bit
  streams: {UTF-8} to {UTF-16} transcoding. In: Proceedings of the 13th ACM
  SIGPLAN Symposium on Principles and practice of parallel programming. pp.
  91--98. ACM (2008)

\bibitem{rfc2781}
Hoffman, P., Yergeau, F.: {UTF-16, an encoding of ISO 10646}.
  \url{https://tools.ietf.org/html/rfc2781} [last checked July 2021] (2000),
  {Internet Engineering Task Force, Request for Comments: 3629}

\bibitem{Hoehrmann}
H\"ohrmann, B.: {Flexible and Economical UTF-8 Decoder}.
  \url{http://bjoern.hoehrmann.de/utf-8/decoder/dfa/} [last checked July 2021]
  (2010)

\bibitem{Inoue2008}
Inoue, H., Komatsu, H., Nakatani, T.: {Accelerating UTF-8 Decoding Using SIMD
  Instructions (in Japanese)}. Information Processing Society of Japan
  Transactions on Programming  \textbf{1}(2), ~1--8 (2008)

\bibitem{keiser2020validating}
Keiser, J., Lemire, D.: Validating utf-8 in less than one instruction per byte.
  Software: Practice and Experience  \textbf{51}(5) (2021)

\bibitem{uci}
{ International Components for Unicode (UCI)}.
  \url{http://site.icu-project.org} [last checked July 2021] (2010)

\bibitem{rfc3629}
Yergeau, F.: {UTF-8, a transformation format of ISO 10646}.
  \url{https://tools.ietf.org/html/rfc3629} [last checked July 2021] (2003),
  {Internet Engineering Task Force, Request for Comments: 3629}

\end{thebibliography}
\end{document}